\documentclass[%
reprint,
superscriptaddress,
amsmath,amssymb,
aps,prl,
]{revtex4-2}

\usepackage{graphicx}
\usepackage{dcolumn}
\usepackage{bm}
\usepackage{siunitx}
\usepackage{amsmath,bm}
\usepackage{amssymb}
\usepackage{textcomp}
\usepackage{color}
\usepackage{braket}
\newcommand{\iu}{\mathrm{i}}

\definecolor{blouge}{rgb}{0.5, .1, .6}
\definecolor{bl}{rgb}{0, .1, .6}
\definecolor{turquoise}{rgb}{0.251, 0.878, 0.816}

\usepackage[commentmarkup=uwave]{changes}
\definechangesauthor[color=blue]{dbo}

\begin{document}

\preprint{APS/123-QED}

\title{Collective early-time spontaneous decay of a strongly driven cold atomic ensemble}

\author{Daniel Benedicto Orenes}
\email[]{daniel.benedicto-orenes@univ-cotedazur.fr}
\affiliation{Universit\'e C\^ote d’Azur, CNRS, Institut de Physique de Nice, 06200 Nice, France}

\author{Naudson Lucas Lopes Matias}
\affiliation{Universit\'e C\^ote d’Azur, CNRS, Institut de Physique de Nice, 06200 Nice, France}

\author{Apoorva Apoorva}
\affiliation{Universit\'e C\^ote d’Azur, CNRS, Institut de Physique de Nice, 06200 Nice, France}

\author{Antoine Glicenstein}
\affiliation{Laboratoire Photonique, Num\'erique et Nanosciences, UMR 5298, CNRS-IOGS-Universit\'e Bordeaux, 33400 Talence, France}

\author{Rapha\"el Saint-Jalm}
\affiliation{Universit\'e C\^ote d’Azur, CNRS, Institut de Physique de Nice, 06200 Nice, France}

\author{Robin Kaiser}
\email[]{robin.kaiser@inphyni.cnrs.fr}
\affiliation{Universit\'e C\^ote d’Azur, CNRS, Institut de Physique de Nice, 06200 Nice, France}


\begin{abstract}
In this work we present a numerical and experimental investigation of the collective early-time decay rates of a strongly driven and optically dense cold atomic cloud. We prepare the atomic ensemble by driving the system to its steady state with varying Rabi frequencies $\Omega$ that go from the weak $\Omega \ll \Gamma$ to the strong driving regime $\Omega \gg \Gamma$, where $\Gamma$ is the single-atom decay rate. We investigate the early-time dynamics in the transition between the strong and weak driving regimes using: i) angular-dependent observables such as the light emitted by the cloud, and ii) global observables, i.e., the excited state population. When driving the cloud on-resonance, we find that as a function of the driving frequency, the behavior of the collected light at certain angles transitions from the single-photon subradiant regime to a superradiant regime while the behavior of the excited state population does not show superradiance. The experiment shows good agreement with numerical predictions in the regime of parameters under study.
\end{abstract}

\maketitle

\paragraph*{Introduction \textemdash} Superradiance, i.e., the collective emission of photons from a medium at a rate $\Gamma_N$ larger than the natural emission rate $\Gamma$ of the emitters composing the medium can have different origins, but it implies some form of coherence within the atomic medium, and it has been observed in a range of physical systems \cite{Tre22,Tim12, Bra17,Ara16, Wan07}.  Applications of superradiant states can be relevant in quantum sensing \cite{Wan14,Pau19,Gie22,Hot25} and quantum metrology, e.g., in superradiant lasers \cite{Boh12,Nad25,Gre12} which have been realized in the last years and are a promising research avenue \cite{Dub25,Nor16,Nor18}, and generally for quantum technology applications \cite{Pen22,Wan20,Gue20,Oli14}. Superradiant phases also appear in many of the most generalized quantum optics models of atom-light interactions \cite{Lar17}, and the study of the transition to such superradiant phases emerges in a variety of physical systems from quantum emitters in solid state systems \cite{Gue20} to optomechanical settings \cite{Jag19}, to ordered arrays of ultra-cold atoms \cite{Mas24}.

\begin{figure}[hb]{\includegraphics[width=.48\textwidth]{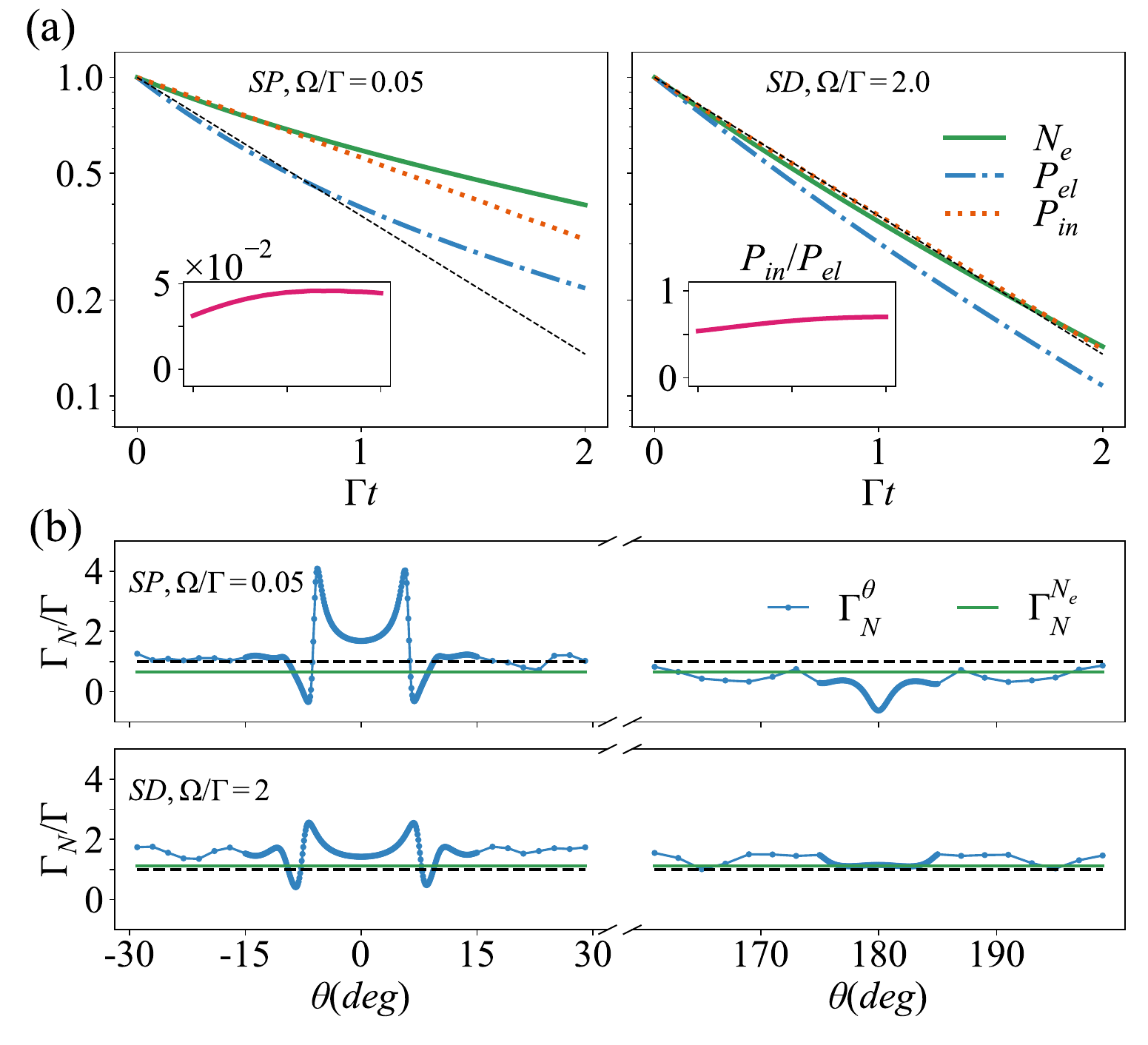}}
\caption{Early-time dynamics from different initial conditions. (a) Illustration of time dynamics of different global quantities normalized to their steady-state values: excited-state population, and total elastic/inelastic emitted power by the cloud. The dynamics is computed using the mean-field model described in the text after driving the system to a steady state with different Rabi frequencies $\Omega/\Gamma = \{0.05,2\}$. The black dashed line shows the single-atom decay for reference. The total elastically emitted power is defined as $P_{el} = \int\int_S I_{el} (\theta,\phi) dS$ where $S$ is the unit sphere. The insets show the time evolution of the ratio between the inelastic and elastic total emitted power in each situation.
(b) Extracted early-time decay rate of the elastically scattered light $\Gamma_N^{\theta}$ (blue points) and $\Gamma_N^{N_e}$ (solid green line) as a function of the detection angle, focusing on the forward and backward directions. The dashed, black line marks the single atom decay rate. In this case, the time interval used for the fitting is $T_{fit} \Gamma = 0.1$. We can see that both in the SP and SD regimes there is a rich angular dependence for the scattered light, which is specially evident in the forward and backward directions, where the influence of the phase induced by the driving field is stronger. Simulations were done with $N = 10^3$ and an optical depth of $b_0 = 8$.}
\label{fig:fig1}
\end{figure}

In the paradigmatic Dicke model \cite{Dic54} the emitters are closely packed within a spatial extent such that $\langle r_{i,j} \rangle \ll \lambda$, and the superradiance effect stems from the fact that  all the dipoles in the system are in phase. The study of Dicke superradiance in atomic systems initiated in the 70s with a series of experimental works reporting the decay dynamics of fully inverted systems in molecular and atomic samples \cite{Skr73,Oka78,Cru78,Bre79,Gro79}. In these works the decay dynamics and the coherence build-up during this process was investigated in detail. This corresponds to studying the decay along the fully symmetric states, the so-called Dicke ladder \cite{Hol25,Ruo25,Fer23,Rob21}. Let us note here that in those works, the state preparation was initiated by indirect optical pumping into the excited state. 

\noindent In cold atomic gases, early-time superradiance has been observed in dilute samples ($\langle r_{i,j} \rangle \gg \lambda$) with high optical depth. During the last decade, due to advances in the theoretical description of collective emission \cite{Scu06,Scu09,Svi13,Svi16}, the experimental research focused on the so-called `single-photon' (SP) sub- and superradiance, i.e., when the driving field is weak enough such that one can consider that the atomic cloud supports a single excitation \cite{Ara16,Roo16,kur17,Gue23}. In this case, the state preparation consists on a weak driving beam of monochromatic laser light which brings the system into its steady state. In this context, superradiant emission properties have been investigated together with its relation with cooperative shifts \cite{Roo16}, in multilevel systems \cite{Sut17}, and also in connection with transport properties in cold atomic gases \cite{Ger06}. The interpretation of early-time sub- and superradiance in cold atomic clouds in the SP regime is well understood in terms of a linear dispersion theory \cite{Wei21,Ass22}. This description breaks down the strong driving (SD) regime, which has not been studied systematically in large ensembles of cold atoms, and is the focus of this work.

In this letter we present a numerical and experimental study of the early-time decay dynamics (after switching off the driving field) of a driven and optically dense cold atomic system composed of two-level atoms, where the collective emission properties stem from the long-range dipole-dipole interactions mediated by the driving field. Up to now, only moderately strong driving fields where investigated either theoretically \cite{Ost24}, or experimentally to test the validity of the mean-field approximation during the switch-on \cite{Esp20} and switch-off \cite{Cip21,Gli25} dynamics, or it was partially explored in dilute samples \cite{Das20}. In this work, we extend the range of study to much stronger driving fields. In terms of state preparation, we drive the system to its steady state while varying the driving intensity characterized by the Rabi frequency $\Omega$. Then we study the early-time dynamics of different observables: i) angular-dependent emitted light from the atomic cloud, and ii) the population of the excited state using an ancillary state \cite{Gli24}. These observables contain different and complementary information about the system \cite{Max20,Gli25}, and we explore the decay dynamics in the transition from the SP to the SD regime. 

\paragraph*{System and model \textemdash} To extend the study of collective atom-light interactions beyond the SP regime studied in previous works \cite{Wei18,Ara16,Gue16} we use a model of coupled dipoles within a mean field approximation, following \cite{Max20,Esp20,Jen18}. The model considers $N$ two-level atoms at positions $\vec{r}_i$ with $i = 1...N$. The atomic transition, with angular frequency $\omega_0$ is characterized by the transition linewidth $\Gamma$. The driving field is monochromatic laser light at frequency $\omega_l$, corresponding detuning $\Delta = \omega_l - \omega_0$, and homogeneous Rabi frequency $\Omega/\Gamma = \sqrt{ s/2 + ( \Delta/\Gamma )^2 }$, where $s = I/I_{sat}$ is the saturation parameter. Neglecting quantum correlations between the atoms to reduce the numerical complexity of the problem, the system can be described by a set of $2N$ non-linear coupled equations where the intensity at each atom has a contribution from the driving laser and a mean-field contribution from every other scatterer in the sample: 

\begin{eqnarray}
    \dot{\beta}_n &=& (\iu \Delta - \frac{\Gamma}{2}) \beta_n + \iu W_n z_n, \label{eq:model_1}\\
    \dot{z}_n &=& - \Gamma ( 1 + z_n) - 4 \mathrm{Im}\left( \beta_n^* W_n \right), 
    \label{eq:model_2}
\end{eqnarray}
with 
\begin{eqnarray}
    W_n &=& \Omega/2 - \iu \sum_{m \neq n} G_{m,n} \beta_m,
    \label{eq:coupling}
\end{eqnarray}
\noindent where $\beta_n = \langle  \sigma_n^- \rangle$ is the coherence of atom at position $\vec{r}_n$, and $z_n = \langle \sigma_n^+ \sigma_n^- - \sigma_n^- \sigma_n^+ \rangle = 2 \langle \sigma_n^z \rangle$. In the scalar approximation, i.e., neglecting near field effects and considering a two-level system, the interaction term is given by $G_{i,j} = (\Gamma \exp[\iu k_l r_{i,j}])/(2\iu k_l r_{i,j}) \mbox{ for }i \neq j$, 
where $k_l = \omega_l/c$ is the wavevector of the laser and $r_{i,j} = |\vec{r}_i - \vec{r}_j|$ the distance between two particles in the sample. This model allows for the calculation of the excited state population 
\begin{equation}
    N_e(t) = \sum_n (z_n(t) + 1)/2, 
    \label{eq:population}
\end{equation}
as well as the elastically scattered intensity by the cloud at a given angle with respect to the propagation axis $\hat{k}_l$ of the driving field 
\begin{equation}
    I_{el}(\theta,\phi,t) \propto  \sum_{i,j} \beta_i^*(t) \beta_j(t) e^{\iu k_l \hat{n} \cdot \vec{r}_{i,j}},
    \label{eq:elastic_intensity}
\end{equation}
where $\hat{n} = \left(\sin\theta \cos \phi, \sin \theta \sin \phi, \cos \theta \right)$ is the unit vector pointing towards the detection area. We also consider the quantity $\Lambda(t) = \sum_i |\beta_i(t)|^2$, which under the assumptions of this model ($\langle \sigma_i^+ \sigma_j^- \rangle \approx \langle \sigma_i^+ \rangle \langle \sigma_j^- \rangle$, for $i \neq j$) can be linked to the total inelastically radiated power \cite{Max20}\footnote{In the weak driving regime, in the limit $\Omega \rightarrow{0}$, $\Lambda$ exactly corresponds to the population of the excited state.}:
\begin{equation}
    P_{in} = \sum_{i = 1}^N \frac{ 1 +  z_i }{ 2 } - |\beta_i|^2.
    \label{eq:inelastic_power}
\end{equation}

\noindent Using the numerical model presented above (Eq.~(\ref{eq:model_1}) and (\ref{eq:model_2})) we calculate the dynamics of $N$ two-level atoms randomly distributed in the three spatial dimensions following a gaussian distribution of standard deviation $R$, and characterized by a peak optical depth $b_0 = 3N / (k_l R)^2$. First, we compute the driving dynamics for a fixed amount of time such that $t\Gamma =10$, and then we switch-off the driving field and compute the decay dynamics. To characterize the early-time decay dynamics we extract decay rates from a fitting procedure of the calculated quantities normalized to their steady state values to an exponential function with a single fit parameter $f(t,\Gamma_N^{fit}) = \exp(-t \Gamma_N^{fit})$, and in a fixed time interval $T_{fit}$.
We note that the model above (Eq.~(\ref{eq:model_1}) and (\ref{eq:model_2})) neglects several phenomena which might contribute when leaving the SP limit. First, the coupling term $W_n$ in Eq.~\ref{eq:coupling} only takes into account the elastically scattered light. The inelastically scattered light, even if with a zero average field, can affect both coherences and populations of other atoms. Second, an initially fully inverted system, with no average atomic dipole moment, is known to lead to superradiant emission after some short delay \cite{Dic54,Skr73,Gro82,rub22}, a feature not captured by Eq.~(\ref{eq:model_1}) and (\ref{eq:model_2}).

\noindent Fig.~\ref{fig:fig1} shows two illustrative examples of the early-time dynamics for different initial states: a cloud prepared in the SP regime, and a cloud prepared in the SD limit. In Fig.\ref{fig:fig1}(a) we show the early-time evolution after the driving field is switched off of different global quantities of the system, which can show sub- and superradiance depending on the initially prepared state. In Fig.\ref{fig:fig1}(b) we show the angular dependence of the extracted decay rate for the scattered light $\Gamma_N^{\theta}$ \cite{kur17} together with the decay rate of the excited state population for reference. This illustrates two things: i) different observables contain different information about the system, both in a quantitative and qualitative manner; ii) although the dynamics of the system in the SD regime tends to the single-atom limit, some observables still show important collective effects.

\begin{figure}[!ht]
{\includegraphics[width=0.45\textwidth]{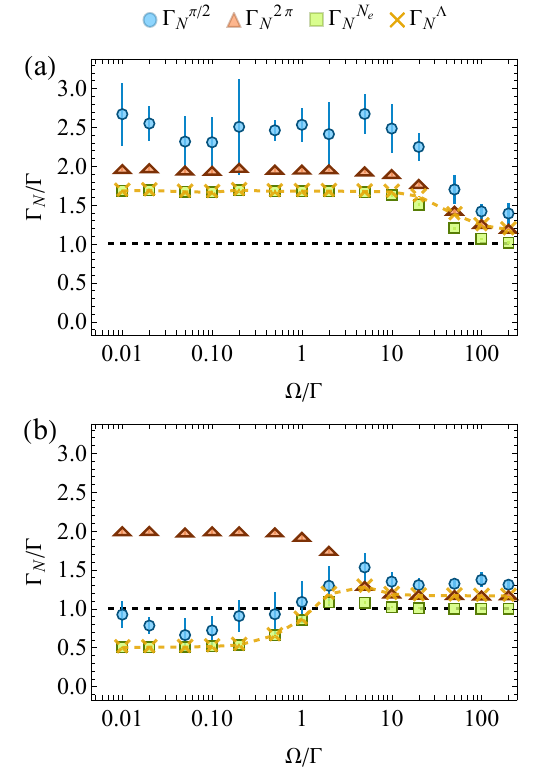}}
\caption{Numerical results on early-time collective decay as a function of on-resonance driving frequency for a cloud with peak optical depth $b_0 = 12$. The black, dashed line marks the single atom behavior. Blue circles (red triangles) are the extracted decay of the light emitted at an angle $\theta = \pi/\SI{2}{\radian}$ ($\theta = 2\pi~\SI{}{\radian}$). Green squares are the results from the decay of the excited state population. Yellow crosses correspond to treat the atomic coherences $\Lambda$ as an observable. The dashed line joining them is just a guide for the eye. (a) Off-resonance driving scenario with $\Delta = -10 \Gamma$. (b) On-resonance driving scenario. For all the simulations we used $N = 5\times10^3$ atoms, and a time $T_{fit} \Gamma = 0.75$ was used to extract the decay rates. We checked that the results do not change in a significant way considering shorter times up to $T_{fit} \Gamma = 0.1$.}
\label{fig:theory}
\end{figure}

\paragraph*{Numerical results and discussion \textemdash}Fig.~\ref{fig:theory} shows the dependence with the driving field intensity of the extracted early-time collective decay rate of different system observables in two scenarios: 
(a) off-resonance driving frequency; (b) on-resonance driving frequency. 

Off-resonant driving:\hspace{7pt}Fig.~\ref{fig:theory}(a) shows the results for the off-resonance driving at a relatively high optical depth $b_0=12$. In this case we observe that superradiance is present for all system observables considered here and all investigated driving frequencies. We do observe how the superradiant behavior is suppressed in the SD limit, specially for the population observable, whose decay rate tends to that of independent atoms. If we look at the behavior of the atomic coherences $\Lambda$, We can notice the transition of $\Gamma_N^\Lambda$ from matching the decay rate of the excited state population to behave as the scattered light in the forward direction. In the low intensity driving limit $\Omega/\Gamma \ll 1$, almost all the incoming light is elastically scattered. However, as the driving intensity increases, the system saturates and the ratio of elastic to inelastic scattering decreases, and the majority of elastically scattered light is then in the forward direction. This is a characteristic that can not be captured in the linear models used to study the SP regime.

On-resonance driving:\hspace{7pt}Fig.~\ref{fig:theory}(b) shows the results of the on-resonance driving, in which the system shows a more complex behavior, both quantitative and qualitatively different for different observables. Interestingly, the model predicts a transition of the decay rate of the off-axis scattered light from a subradiant to a superradiant regime as a function of the driving intensity.
To understand this effect we can reason from the SP limit, where the emission rate of the off-axis scattered light is subradiant when the system is driven by light on-resonance with the atomic transition. This is due to the enhancement of multiple scattering processes within the medium \cite{Ara16}. Increasing the intensity of the driving field saturates the atomic medium, and the model shows that a superradiant regime is recovered. The decay rate of the excited state population is also significantly affected by the driving intensity and goes from a markedly subradiant behavior in the weak excitation limit $\Omega/\Gamma \ll 1$ to a slightly superradiant one, and tends to the limit of independent atoms at high driving frequencies. Again, the behavior of $\Gamma_N^\Lambda$ is to follow $\Gamma_N^{N_e}$ in the weak driving scenario and then transitions to match $\Gamma_N^{2 \pi}$ in the SD case due to the weighting of the elastic to inelastic scattering. The emitted light in the forward direction shows a very similar behavior as the one in the off-resonance scenario.

Summarizing, we can see that the numerical model predicts less pronounced collective effects in the SD limit, and that it shows an interesting transition from subradiance to superradiance of the off-axis emitted light measurements after the cloud is prepared by an on-resonance driving field at moderate intensities $\Omega \sim \Gamma$.

\begin{figure}[ht]
\includegraphics[width=0.45\textwidth]{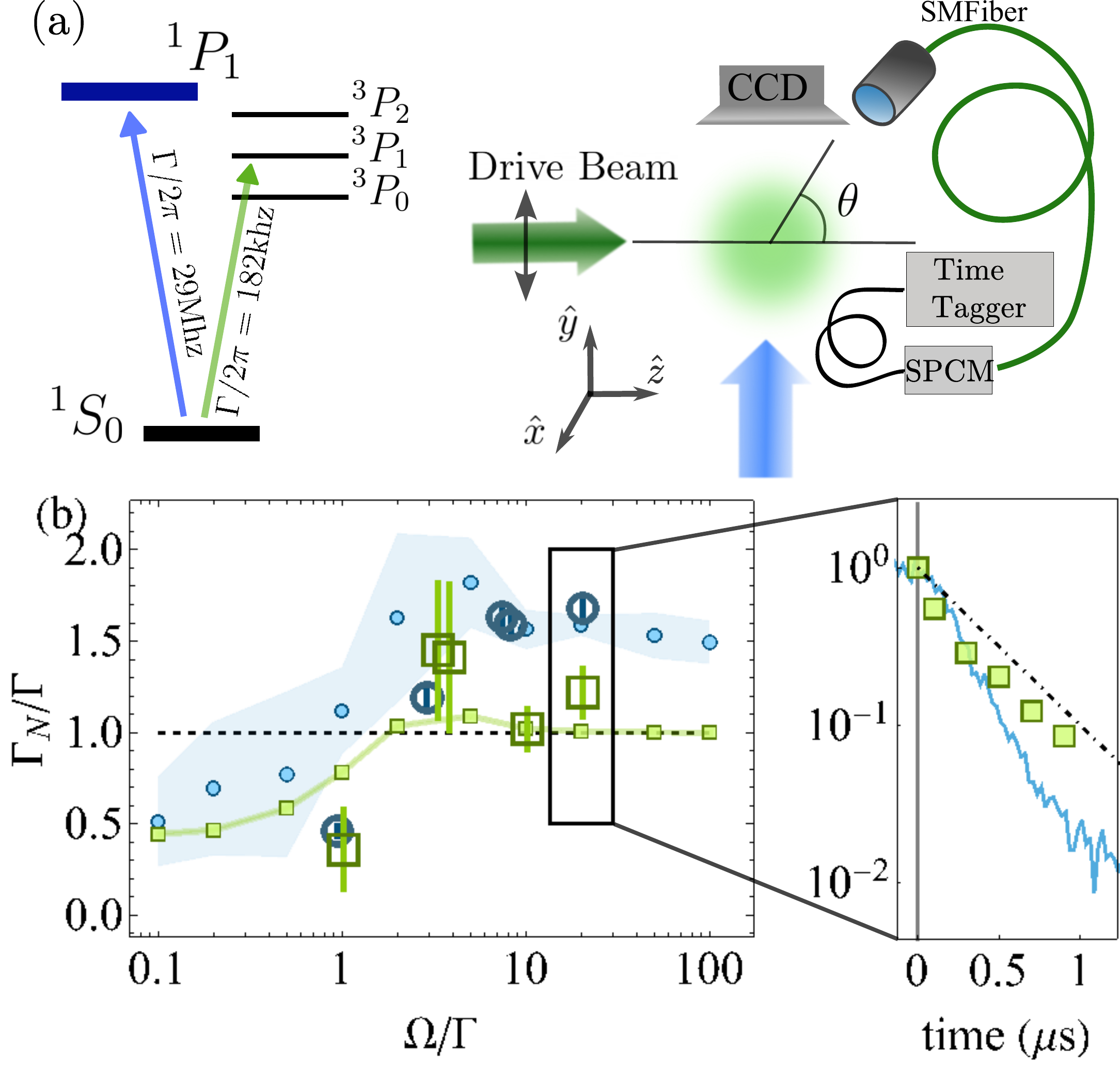}
\caption{Experimental results. (a) Energy level scheme of $^{174}$Yb atoms and schematic representation of the experimental setup described in the main text. (b) Measured early-time decay rates together with numerical predictions for the elastically scattered light at $\theta = \pi/3$ (blue points) and excited state population (green squares). Small full markers correspond to numerical simulations. For these simulations, we used $b_0 = 15$ and $N = 8\times 10^3$. The confidence bands on the numerical results correspond to the standard deviation of 5 different random atomic configurations. Big empty markers corresponds to experimental results. The error bars on the experimental points represent the uncertainty in the fitting procedure to extract the exponential decay constant. For both the theoretical and the experimental data we used a fitting time $T_{fit}\Gamma \approx 1$. The inset shows an example of a time trace of the decay dynamics of the system as recorded using the intensity (blue trace) and population (green squares) methods for the same driving conditions. Both signals are normalized to their steady-state values. The dot-dashed line corresponds to the single atom decay dynamics.}
\label{fig:experiment}
\end{figure}

\paragraph{Experimental setup and detection \textemdash} The experimental setup is detailed in previous works \cite{Letellier2023,deMelo2024,Gli24} and is represented in Fig.~\ref{fig:experiment}(a). First, a magneto-optical trap (MOT) operating on the ${}^{1}S_0 - {}^{1}P_1$ transition of $^{174}\mathrm{Yb}$ ($\lambda_{\mathrm{b}}=\SI{399}{\nano\meter}$, $\Gamma_{\mathrm{b}}=2\pi\times\SI{29}{\mega\hertz}$) is loaded in $\sim \SI{1}{\second}$. Then, about $10^8$ atoms are transferred into a second MOT on the ${}^{1}S_0 - {}^{3}P_1$ intercombination line ($\lambda_{\mathrm{g}} = \SI{555.8}{\nano\meter}$, $\Gamma_\mathrm{g} = 2\pi\times\SI{182}{\kilo\hertz}$). The resulting cloud has a final temperature of $\sim \SI{15}{\micro\kelvin}$ and typical size $R_{0} \approx \SI{400}{\micro\meter}$. Then the magnetic field and optical beams are switched off, and the atomic cloud expands in free fall for a fixed time-of-flight (TOF). We adjust this TOF such that the cloud has a peak optical depth $b_0 \approx 12-16$. A linearly polarized probe beam with Rabi frequency $\Omega$, waist $w \approx \SI{1200}{\micro\meter}$ and resonant with the $^1S_0 \rightarrow{} ^3P_1$ transition is switched on during $\SI{10}{\micro\second} \gg 1/\Gamma_{\mathrm{g}}$ to ensure that a steady state is reached. Then, it is switched off, and we extract information about the decay dynamics of the system using two different observables: total (i.e., sum of elastic and inelastic) scattered light collected at an angle $\theta$, and the excited state population. These methods have been described in detail in previous works \cite{Gli24,Gli25}, and we just recall here their basic features. For the measurement of the scattered intensity, we collect the scattered photons that impinge in a solid angle $\simeq \SI{0.01}{\steradian}$ and are coupled into a single-mode fiber placed at an angle $\theta = \pi/\SI{3}{\radian}$ with respect to the propagation axis of the probe beam. The fiber is connected to a single photon counting module (SPCM-AQRH-14-FC, Excellitas) whose output is connected to a counting card (Time Tagger Ultra, Swabian Instruments). Scattered photons are recorded during $\SI{20}{\micro \second}$ corresponding to the initial $\SI{10}{\micro\second}$ of driving plus another $\SI{10}{\micro\second}$ of decay after we switch off of the excitation pulse. To increase the collected intensity, we send up to 20 pulses of probe light per experimental realization. We integrate over many ($\sim 2 \times 10^4$) experimental repetitions until sufficient signal is collected, depending on the experimental conditions. For the measurement of the excited state population, we use the imaging technique demonstrated in \cite{Gli24}. We  measure the dynamics of the excited state population by taking absorption images using a beam resonant with the $^1S_0 - ^1P_1$ transition. First, we prepare an atomic sample and take a reference image $I_{\mathrm{r}}$ with no excited state population in the $^3P_1$ state. Then, we prepare another atomic sample, we drive the system as described above, and we allow the system to decay for a given time before we take the depleted image $I_{\mathrm{d}}$. We take $\sim 50$ images for averaging. Within the regime of parameters that we use in this work~\cite{Gli25}, we can extract the excited state population from the ratio between the two images $N_e \approx \sigma_{\mathrm{sc}}^{-1} \int{\int{ R(x,y)}{\ \mathrm{d}x \ \mathrm{d}y }}$, where $R(x,y)=I_{\mathrm{d}}(x,y)/I_{\mathrm{r}}(x,y)-1$, and $\sigma_{\mathrm{sc}}$ is the scattering cross section of the atoms with respect to the imaging transition $^1S_0 - ^1P_1$. The duration and delay of the probe and imaging pulses are synchronized using a delay generator, and the imaging pulse is set to $\SI{150}{\nano\second} \ll 1/\Gamma_g$ such that during the imaging we can consider the dynamics in the $^1S_0 - ^3P_1$ probing transition is frozen. Regarding the experimental method, we note that the use of a narrow transition allows us to easily access the early-time dynamics of the system with typical optical modulators and detectors. Also,  since all the Clebsch-Gordan coefficients of the transition are equal to one, and we compensate external magnetic fields, all the energy levels are degenerate, offering a good analog of a true two-level system.

\paragraph*{Discussion of experimental results \textemdash}Fig.~\ref{fig:experiment}(b) summarizes the experimental results for the on-resonance driving scenario. We plot the extracted early-time decay rates $\Gamma_N^{\theta_{\pi/3}}, \Gamma_N^{N_e}$ together with the results from numerical simulations for equivalent optical depths, detection angle, and driving frequencies. We can appreciate the transition from the subradiant to superradiant regime as a function of the driving frequency for the scattered intensity, as discussed previously. Also, we observe that the dynamics of the excited state population starts well in the subradiant regime for $\Omega/\Gamma \ll 1$, and tends to the single atom decay at larger driving frequencies. In general, quantitative agreement is good for $\Omega/\Gamma > 1$, while in the intermediate driving regime $\Omega \sim \Gamma$, we still find good qualitative agreement with the numerical predictions. We attribute the deviation of the experimental results from the theoretical predictions in this regime to the use of a gaussian intensity distribution with size comparable to the atomic cloud to drive the system, while the numerical results assume a plane wave. This, in combination with the very small saturation intensity is $I_s = \SI{0.14}{\milli\watt / \centi\meter^2}$ of the narrow transition used in this work (thus the very small intensity $\approx \SI{15}{\micro\watt}$ used in the probe beam to achieve the lowest driving frequencies $\Omega/\Gamma \sim 1$) make the results more sensitive to experimental imperfections, e.g., thermal effects of the acousto-optical modulators during the switching-on dynamics of the probe beam. The inset of Fig.~\ref{fig:experiment}(b) shows two particular cases of the measured dynamics for the two observables considered at the same driving intensity, showing a markedly different behavior, as predicted by the theory.

\paragraph{Conclusions \textemdash} In this work, we used the non-linear coupled-dipoles model to explore numerically the behavior of a system of two-level atoms in free space in the SD regime. The model predicts a non-trivial behavior for the off-axis scattered light when driving the system on-resonance. We experimentally confirm these results, showing that the dynamics of the off-axis scattered light undergoes a transition from a suppressed superradiant regime into a superradiant regime as a function of the driving intensity. In the case of the excited state population, the behavior tends to the limit of independent emitters in the SD limit. These `a priori' discordant results illustrate the importance of having access to different system observables, and shed light into the behavior of optically dense media under strong-driving. In this situation, a complex dynamics and a rich interplay between elastic/inelastic scattering takes place during the driving, which in turn affects the posterior decay, highlighting the importance of initial state preparation for the decay dynamics. In summary, we have explored a novel set of initial state parameters to study the early-time decay dynamics of an optically dense, isotropic and diluted cold atomic cloud, validating the non-linear mean-field model used to describe the system in this regime. This serves as a first benchmark for further studies of collective behavior in the SD regime, where atom-atom correlations and beyond-mean-field effects are expected to be present \cite{Cip21,Rob21} due to the strong light mediated dipole-dipole interactions. Some possible directions for future work include the measurement of the light scattered around the forward direction, possibly with angular resolution, the measurement of second-order coherence functions to better understand the nature of the induced sub- and superradiance, or exploring partially and quasi-fully inverted systems prepared by direct driving of the atomic transition \cite{Lie23}. These experiences will help, for instance, to increase our understanding of atom-light interactions in open quantum systems, a regime where classical computational power limits the number of simulated scatterers or analytical results are available only under important geometrical assumptions \cite{Hol25}, but also to design better quantum information protocols with potential applications in modern quantum technologies \cite{Hou24,Ras22} and fundamental research \cite{Loh23,Kar21}. 

This work was performed in the framework of the European project ANDLICA (ERC Advanced grant No. 832219), the French National Research Agency (projects PACE-IN (ANR19-QUAN-003), LiLoA (ANR23-CE30-0035) and QUTISYM (ANR-23-PETQ-0002)). D.B.O is supported by European Union’s Horizon 2020 research and
innovation program under the Marie Skłodowska-Curie grant agreement No. 10110529.

\begin{acknowledgments}

\end{acknowledgments}

\bibliography{apssamp}

\end{document}